\documentclass[sigchi-a]{acmart}
\usepackage{booktabs} 
\usepackage{ccicons}  

\setcopyright{acmcopyright}

\acmDOI{10.475/123_4}

\acmISBN{123-4567-24-567/08/06}

\acmConference[CHI 2019]{The ACM CHI Conference on Human Factors in Computing Systems}{May, 2019}{Glasgow, UK}
\acmYear{2019}
\copyrightyear{2019}
\acmPrice{15.00}

\copyrightyear{2019}
\acmYear{2019}
\setcopyright{rightsretained}
\acmConference[CHI'19 Extended Abstracts]{CHI Conference on Human Factors in Computing Systems Extended Abstracts}{May 4--9, 2019}{Glasgow, Scotland Uk} \acmBooktitle{CHI Conference on Human Factors in Computing Systems Extended Abstracts (CHI'19 Extended Abstracts), May 4--9, 2019, Glasgow, Scotland Uk}\acmDOI{10.1145/3290607.3312816}
\acmISBN{978-1-4503-5971-9/19/05}

\begin{document}
\title[Usability of VR Application Through the User Community: A Case Study]{Usability of Virtual Reality Application Through the Lens of the User Community: A Case Study}

\author{Wenting Wang}
\affiliation{%
  \institution{McGill University}
  \city{Montreal}
  \state{Quebec Canada}
}
\email{wenting.wang@mail.mcgill.ca}

\author{Jinghui Cheng}
\affiliation{%
  \institution{Polytechnique Montreal}
  \city{Montreal}
  \state{Quebec Canada}
}
\email{jinghui.cheng@polymtl.ca}

\author{Jin L.C. Guo}
\affiliation{%
  \institution{McGill University}
  \city{Montreal}
  \state{Quebec Canada}
}
\email{jin.guo@cs.mcgill.ca}

\renewcommand{\shortauthors}{W. Wang et al.}



\begin{abstract}
The increasing availability and diversity of virtual reality (VR) applications highlighted the importance of their usability. Function-oriented VR applications posed new challenges that are not well studied in the literature. Moreover, user feedback becomes readily available thanks to modern software engineering tools, such as app stores and open source platforms. Using Firefox Reality as a case study, we explored the major types of VR usability issues raised in these platforms. We found that 77\% of usability feedbacks can be mapped to Nielsen's heuristics while few were mappable to VR-specific heuristics. This result indicates that Nielsen's heuristics could potentially help developers address the usability of this VR application in its early development stage. This work paves the road for exploring tools leveraging the community effort to promote the usability of function-oriented VR applications.
\end{abstract}


\maketitle

\section{Introduction}
Virtual reality (VR) is a future-shaping technology that may hugely impact our interaction paradigm with computer systems. With the increasing availability of commercially off-the-shelf hardware devices and the proliferating software development frameworks, VR applications are experiencing rapid growth and started to break into the mass market. This growth is reflected in the increasing number of applications available in VR app stores (e.g. Oculus Experiences) and the boost of open source VR projects (e.g. on GitHub) \cite{Rodriguez2017, Ghrairi2018}.
For users, increased availability of VR applications makes it easier to access the immersive experience to complete various tasks. Naturally, the increasing and diversified user base puts the usability of VR applications to the test \cite{Oliveira2017}.

With the support of modern software engineering tools and platforms, community feedback about usability, especially from the users, is readily available to developers:
On one hand, open source projects bring closer the distance between the users and the developers by offering a rich dynamic environment where users' feedbacks are welcomed and directly communicated with the developer team during the development cycle \cite{Cheng2018}. On the other hand, app store reviews provide a direct channel for users to voice opinions about a VR application after release \cite{Pagano2013}. However, given the special interaction schema of VR, it is not well understood as to what types of information pertinent to improving VR usability were raised in the discussions and feedback on these platforms.

Moreover, previous work on VR usability mainly focused on simulations and games \cite{Kaur1997, Sutcliffe2018}; many focused on heuristic evaluation \cite{Sutcliffe2004, Rusu2011, Oliveira2017}. The modern VR applications, however, has expanded to a diverse landscape, including communication, e-commerce, and productivity tools, to name a few. These function-oriented applications posed new challenges to usability that are not well studied in the literature. The existing VR usability guidelines and heuristics may or may not apply. In these scenarios, the direct feedback from the user community becomes more valuable for developers of diverse VR applications.

To help address these challenges, we investigate the following research questions: 
\begin{itemize}
    \item\textbf{RQ1}: What are the major types of usability issues about function-oriented VR applications raised in open source platforms and app store reviews?
    \item\textbf{RQ2}: To what extent can existing usability heuristics reflect users' opinions on the usability of function-oriented VR applications?
\end{itemize}

In this paper, we explore these RQs through a case study of Firefox Reality, an open source VR web browser developed by Mozilla. We chose this application because: (1) it is a new project initially released in September 2018 that has a dynamic community welcoming to improvement suggestions and (2) it is an application that transforms a habitual user interaction (i.e. web browsing) into the VR world and thus may induce interesting usability concerns. Particularly, we conduct qualitative analysis of its GitHub issue discussions \footnote{GitHub \textit{Issues} provide a space for everyone to report and discuss enhancements, bugs, and tasks concerning the open source project, including its usability \cite{Cheng2018}} and Oculus app store reviews and try to map the usability themes we identified in the user community on these platforms with existing usability heuristics.




\section{Methods}



Data collection happened in November 2018. We used the GitHub REST API to retrieve the GitHub issues and comments and manually fetched all Oculus app store reviews about Firefox Reality. In total, we collected 826 GitHub issue threads and 20 app store reviews.

Upon completing data collection, we performed a qualitative analysis with the following steps. We first cleaned the dataset by removing (1) the issues explicitly marked as a duplicate on GitHub and (2) the issues/reviews with only an imprecise title and no body content. We then labeled the remaining 672 issues/reviews as usability-related or not based on their main content. For the usability-related issues/reviews, we performed open coding to identify the common themes of usability concerns.
Next, we attempted to map all the identified themes to Sutcliffe's \cite{Sutcliffe2004} and Rusu's \cite{Rusu2011} heuristics for virtual environments, as well as Nielsen's \cite{nielsen1995} heuristics for general applications. When mapping the themes to these heuristics, we found that Sutcliffe's and Rusu's heuristics were applicable to very few usability themes we identified. We speculate that this is due to the mismatch of these heuristics (i.e. focused on immersive virtual environments) and the nature of the Firefox Reality application (i.e. mimicking desktop application in a virtual world). Consequently, we only focused our mapping with Nielsen's heuristics. The coding and mapping processes are first performed by one researcher. The research team then met to review the analysis results and discuss any ambiguity and disagreement. The final coding and mapping are formed based on consensus after discussion.

\section{Results}



\begin{marginfigure}
    \includegraphics[width=\marginparwidth]{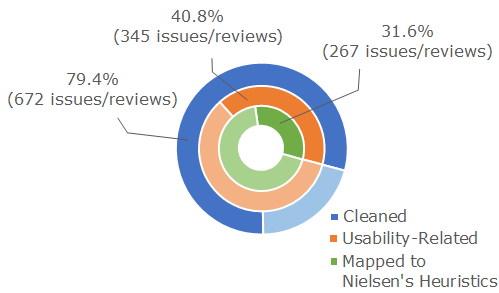}
    \caption{Percentage of Issues/Reviews out of 846 Total Number of Issues/Reviews}
    \label{fig:percentage}
\end{marginfigure}

Within the set of 672 cleaned issues/reviews, we identified 345 that are usability related. The open coding process yielded 14 usability themes. Among all usability-related issues, 267 (77.4\%) were mapped to at least one of Nielsen's heuristics (see Figure \ref{fig:percentage}). Below, we report the mapped heuristics with their examples, followed by the themes that could not be mapped to the heuristics.

\subsection{Nielsen's Heuristics}

\begin{marginfigure}
    \includegraphics[width=\marginparwidth]{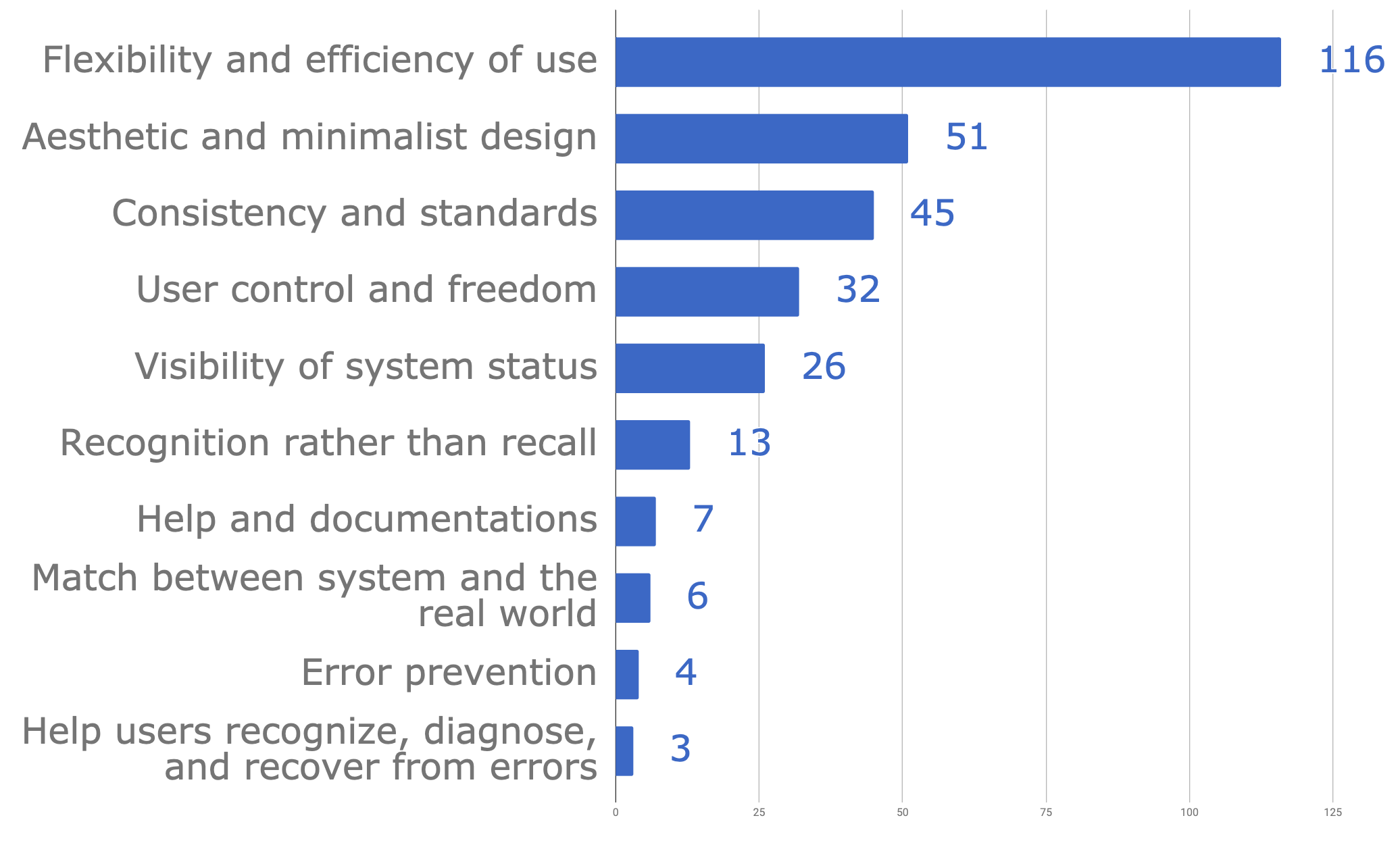}
    \caption{Distribution of issues/reviews mapped to Nielsen's Heuristics}
    \label{fig:distribution}
\end{marginfigure}

Figure \ref{fig:distribution} shows the number of issues/reviews mapped to each of Nielsen's heuristics.

\subsubsection{Flexibility and efficiency of use} 
Firefox Reality users raised concerns for the inefficiency of interaction due to the unconventional input medium provided by the VR system. As issue 775 reporter complained that "the Oculus Go only has a point-and-click device and it becomes annoying to write text with that" and further suggested to add keyboard support, issue 355 described that "users wanted 'voice to text' support to make it easier to enter text strings".

\vspace{-3pt}
\subsubsection{Aesthetic and minimalist design} 
With the aim of delivering a system with simple and aesthetic GUI, the Firefox Reality developer team worked with designers throughout the development cycle and actively iterated the design in response to user feedback. E.g. Issue 281 tracked the need to "change hover style and text selection color for URL bar".

\vspace{-3pt}
\subsubsection{Consistency and standards}
Because Firefox Reality brings a familiar desktop application into the virtual world, users' expectations for this new project were sometimes based on their acquaintances to the traditional product. E.g., reporter of Issue 705 says s/he expected "swappable Y-Axis option for touchpad scrolling" because it is a common feature on traditional web browsers.

\vspace{-3pt}
\subsubsection{User control and freedom}
Users sometimes felt losing control during page navigation. This may be due to product defects or inappropriate design choice. E.g., reporter of Issue 235 expressed desires to "disable zoom on double click" because it caused the problem of "page zoomed in without a way to zoom back out".

\vspace{-3pt}
\subsubsection{Visibility of system status} 
While some users reported missing feedbacks in the traditional way, others also expressed their wishes for the product taking better advantage of the virtual medium. E.g. Issue 353 stated that "users wanted haptic feedback when using controls and navigating."

\vspace{-3pt}
\subsubsection{Recognition rather than recall}
Firefox Reality users demonstrated interests in having functionalities that can help reduce their memory load. These functionalities include auto-filling and restoring sessions upon reopening or crashing. E.g., Issue 91 complains that "sessions are not correctly restored in some situations".

\vspace{-3pt}
\subsubsection{Help and documentation}
In addition to the traditional user guide documents, some users showed that they would appreciate more contextual 'help and documentation' in the VR world, such as 'Learn More' links and tooltips. E.g., Issue 516 stated that developers needed to "add SUMO help link to 'Learn more about Private Browsing' on the Private Browsing splash page".

\vspace{-3pt}
\subsubsection{Match between system and the real world}
The Firefox Reality community also emphasized delivering the system that matches to the real world. Particularly, the look and feel of the background environment were extensively discussed. E.g., Issue 171 stated its goal is to "set up correct lighting for meadow environment". Furthermore, issue 177 and 389 suggested adding a fading effect and "implementing Sunrise/Sunset (mapped to brightening/dimming scene) when entering/exiting Private Mode" to achieve an eye-comfortable and close-to-nature environment.

\vspace{-3pt}
\subsubsection{Error prevention}
Users suggested needs to receive confirmation for error-prone actions to avoid mistakes, which can be easily made in a VR system. E.g., Issue 568 suggested to "prompt users if they want to quit on back button press".

\vspace{-3pt}
\subsubsection{Help users recognize, diagnose, and recover from errors}
The Firefox Reality developer team recognized the importance of providing useful, as well as localized error messages. E.g., Issue 803 expressed the need for "better error message for offline WiFi/Internet connection".


\subsection{Other Themes}
\subsubsection{Security and Privacy Concerns (22 issues/reviews)} 
Firefox Reality introduced telemetry to help the developers understand users' activities. However, this approach is at the cost of rising users' concerns towards security and privacy. E.g., Issue 52 stated that there is a need for "preferences/settings for telemetry, crash reporting, privacy policy." Notably, some of the issues/reviews from this theme may be related to other themes; e.g., the previous example can be mapped to "Flexibility and efficiency of use". However, we coded it separately because it is a sensitive issue brought up by many users.

\subsubsection{Accessibility (9 issues/reviews)} 
The Firefox Reality users and the developer team realized the importance of creating products available to everyone. E.g., Issues 423 and 449 voiced the need for textual labels on the toggle switch in the settings panel in order to accommodate users who are colorblind.

\subsubsection{Camera Control (3 issues/reviews)}
Firefox Reality users wished to adjust their camera view. E.g., Issue 348 stated that "users occasionally felt as though their head position was too high above ground".

\subsubsection{Faithful viewpoint (3 issues/reviews)}
Users also complained that the viewpoint change is sometimes not faithful to head movement. Delay in view rendering sometimes causes motion sickness. A problem of "controller drifts when moving the head" is captured in Issue 185.


\section{Discussion and Conclusion}
In this paper, we conducted a qualitative analysis of GitHub issues and Oculus app store reviews for the Firefox Reality project in order to identify its major types of usability issues reflected in these user feedback channels.
We found a 77\% mapping rate from the usability-related issues/reviews to Nielsen's heuristics, an early yet far-reaching work, indicating that Nielsen's heuristics can sufficiently capture user's opinion on this VR application and can be used for the developers to improve its usability. It is worth noting that Firefox Reality is still at an early stage of development. So the product is described by some users (e.g. Reviewer 9 on Oculus app store) as "nothing more than the default browser." This might explain the high mapping rate.

There are some interesting VR specific findings among the themes mapped to the heuristics. For example, users extensively discussed efficiency loss due to the change of the input medium. As a solution, text dictation is introduced to compensate for the inefficient point-and-click on the virtual keyboard. More interestingly, transiting a traditional website into VR also causes changes in familiar concepts; for example, the term "full screen" becomes ambiguous for applications in VR.

We also identified four unmapped themes. As the use of VR systems become prevalent, accessibility (i.e. making software applications to accommodate users with a disability) becomes an important factor. Security and privacy concerns also raised together with the increasing amount of digital data shared through VR applications. Navigation in VR systems is a long lasting issue and is also exhibited in function-oriented applications such as Firefox Reality. Camera control and faithful viewpoint are the two themes that reflect this aspect. In this respect, usability heuristics specifically identified for virtual environments could be relevant.

As future work, we will first expand this study by examining other function-oriented VR applications to assess the generalizability of our results. With the combination of existing heuristics and further analysis, we aim to investigate usability heuristics and guidelines tailored for function-oriented VR applications. Eventually, we plan to use this knowledge to explore semi-automated tools that categorize user feedback based on usability heuristics and guidelines in order to provide concrete support to developers of function-oriented VR applications.




\section{Acknowledgments}
This work was partially funded by the Canada NSERC Dicovery Grant RGPIN-2018-04470.

\bibliographystyle{ACM-Reference-Format}


\end{document}